# AUTO-CORRELATION FUNCTIONS AND SOLAR CYCLE PREDICTABILITY

*Stefano Sello*
*Mathematical and Physical Models*
*Enel Research*
*Via Andrea Pisano, 120*
*56122 PISA - ITALY*

*ABSTRACT*

Starting from the traditional definition of auto-correlation function we can introduce a new quantity, based on the wavelet formalism, called local wavelet auto-correlation function which contains more information about the correlation properties of a given signal on different scales. The application of this formalism to the temporal variations of solar activity, from the monthly time series of sunspot numbers, allows to derive some interesting information about the correlation properties of different periodicities present in the signal and in particular the correlation characteristics of the main 11 years Schwabe cycle. The scale selectivity, allowed by the local wavelet auto-correlation function, can give interesting information about the range of reliable predictions reachable through deterministic predictive models.

INTRODUCTION

The traditional way to record the variations of solar activity is to observe the sunspot numbers which provide an index of the activity of the whole visible disk of the Sun. These numbers are determined each day and are calculated counting both isolated clusters of sunspots, or sunspot groups, and distinct spots. The current determination of the international sunspot number, Ri, from the Sunspot Index Data Center of Bruxelles, results from a statistical elaboration of the data deriving from an international network of more than twenty five observing stations. Despite sunspot numbers are an indirect measure of the actual physics of the Sun photosphere, they have the great advantage of covering various centuries of time. This property results very crucial when we focus the attention on complex variations of scales and cycles related to the solar magnetic activity.
Many previous works based on standard correlation and Fourier spectral analyses of the sunspot numbers time series, reveal a high energy content corresponding to the Schawbe (~11 years) cycle. More careful statistical analyses show that this periodicity is in fact strongly changing both in amplitude and in scale, showing clear features of transient phenomena and intermittency. Wavelet analysis offers an alternative to Fourier based time series analysis, especially when spectral features are time dependent. By decomposing a time series into time-frequency space we are able to determine both the dominant modes and how they vary with time. Multiscaling and intermittency features have been well analysed through the use of a continuous Morlet wavelet transform on various sunspot numbers time series [1], [2], [3].

The non-linear and chaotic nature of the dynamics underlying the sunspot time series has been suggested and well evidenced by many authors [4],[5],[6]. This characteristic forced a strong limitation on the predictability range of deterministic forecasting methods. The most accurate and refined non-linear analyses show that the typical predictability time is about 4 years, suggesting a little reliability for long term forecasting of solar activity [5],[6].

A traditional way to analyse the time correlation properties of a given signal is the computation of the auto-correlation functions, which describe the general linear dependence of the values of signal at a given time on the values at another lagged time. The related information is global, i.e. an average evaluation of the time dependences over all the scales. On the other hand, using the wavelet formalism, it is possible to extend the auto-correlation analysis in order to obtain a local, scale dependent, information about the correlation of the signal.

WAVELET AUTO-CORRELATION FUNCTIONS

The continuous wavelet transform represents an optimal localized decomposition of time series, $f(t)$, as a function of both time $t$ and frequency (scale) $a$, from a convolution integral:

$$W_f(a,\tau) = \frac{1}{a^{1/2}} \int_{-\infty}^{+\infty} dt\, f(t)\psi^*\left(\frac{t-\tau}{a}\right)$$

where $\psi$ is called an analysing wavelet if it verifies the following admissibility condition:

$$c_\psi = \int_0^{+\infty} d\omega\, \omega^{-1} \left|\hat{\psi}(\omega)\right|^2 < \infty$$

where:

$$\hat{\psi}(\omega) = \int_{-\infty}^{+\infty} dt \, \psi(t) e^{-i\omega t}$$

is the related Fourier transform. In the definition, *a* and *τ* denote the dilation (scale factor) and translation (time shift parameter), respectively.

Here we use the family of complex analysing wavelets consisting of a plane wave modulated by a Gaussian (called Morlet wavelet) [7]:

$$\psi(\eta) = \pi^{-1/4} e^{i\omega_0 \eta} e^{-\eta^2/2}$$

where $\omega_0$ is the non dimensional frequency here taken to be equal to 6 in order to satisfy the admissibility condition. For a more comprehensive and detailed description of the wavelet formalism see references [7], [8].

The standard auto-correlation function is calculated as [9]:

$$C(\tau) = \lim_{T \to \infty} \frac{1}{T} \int_{-T/2}^{T/2} dt f(t) f(t+\tau)$$

where f(t) here is the monthly sunspot number time series. In general we prefer the normalized version:

$$R(\tau) = \frac{C(\tau)}{C(\tau=0)}$$

In the wavelet formalism we can generalise the concept of the auto-correlation function if instead of calculating the correlation of the signal f(t) directly we calculate the auto-correlation of the wavelet transform of f(t). Following Forsth [10] we define the *wavelet auto-correlation function* as:

$$WC(a,\tau) = \lim_{T \to \infty} \frac{1}{T} \int_{-T/2}^{T/2} dt \, W_f^*(a,t) W_f(a,t+\tau)$$

Because WC(a,τ) is a complex function we prefer, for simplicity, to consider its real part, which also contains information about the sign of the auto-correlation:

$$RWC(a,\tau) = \Re(WC(a,\tau)) =$$
$$= \lim_{T \to \infty} \frac{1}{T} \int_{-T/2}^{T/2} dt\, \Re(W_f(a,\tau))\Re(W_f(a,t+\tau)) +$$
$$+ \Im(W_f(a,\tau))\Im(W_f(a,t+\tau))$$

and the normalized version is called the wavelet auto-correlation coefficient:

$$RWR(a,\tau) = \frac{RWC(a,\tau)}{C(\tau=0)}$$

Here we prefer to use a local normalization, i.e. a normalisation different for each scale a:

$$LRWR(a,\tau) = \frac{RWC(a,\tau)}{RWC(a,\tau=0)}$$

The function LRWR(a,τ) is called *local wavelet auto-correlation function*.

The usefulness of the wavelet auto-correlation functions, to investigate multiscaling unsteady processes, has been recently well documented by Forsth on applications to turbulent velocity fields [10].

SOLAR ACTIVITY: SUNSPOT NUMBERS

As a record of solar activity we considered here the monthly international number of sunspots from SIDC archive covering the time interval: 1749-8/2000 and consisting of 3020 data [11] (see Figure 1). Previous detailed wavelet analyses showed the significant periodicities inside this time series and their complex time evolution [2], [3]. Here we are mainly interested in the correlation properties of the strong Schwabe cycle of about 11 years. This cycle is in fact the most relevant when we consider the reliability of solar cycle predictive methods.

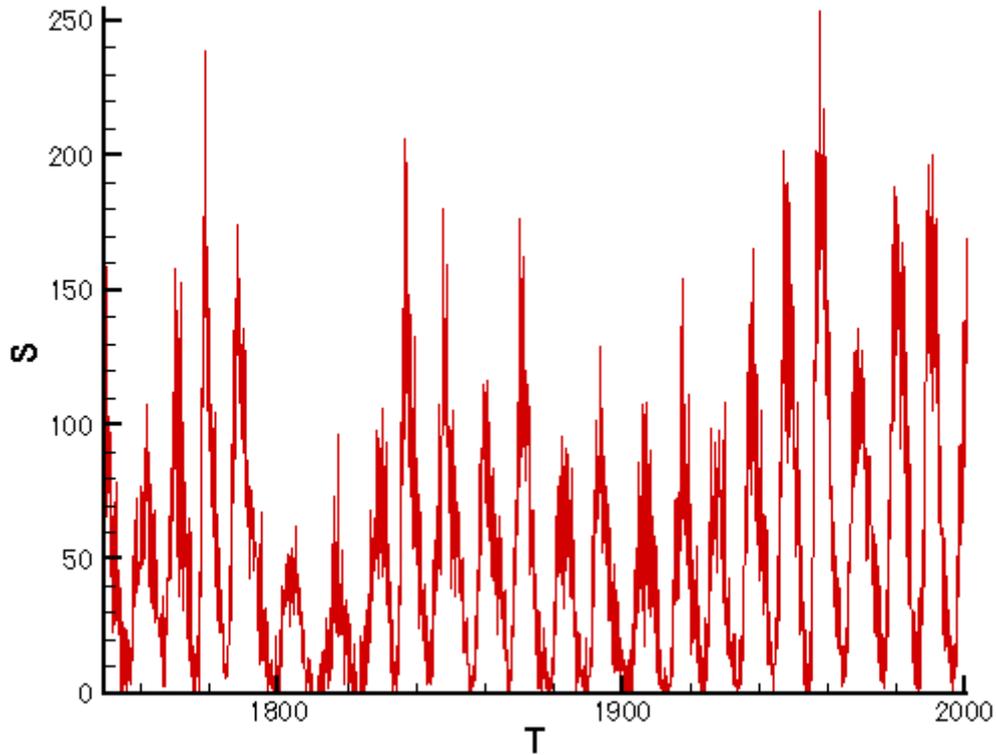

Figure 1 - Monthly sunspot numbers from SIDC

In the upper part of Figure 2 we show the computation of the standard auto-correlation function applied to the sunspot time series. Time is here expressed in years. By comparison in the lower part we show the local wavelet auto-correlation function in terms of an arbitrary colour contour map. Red and blue higher values are strong absolute correlation contributions, while green lower values are weak correlation contributions. Horizontal time axis corresponds to the time lag in years and vertical scale (frequency) axis is for convenience expressed in log values of cycles per year$^{-1}$. Thus the scale range analyzed is between 54 years (value -4) and 134 days (value 1).

From the standard auto-correlation function we can derive an average information about the decay of linear dependence or correlation inside the data. The amplitude of oscillations of the auto-correlation function remains significant up to about 50 years, but its value decays to zero at about 3 years.

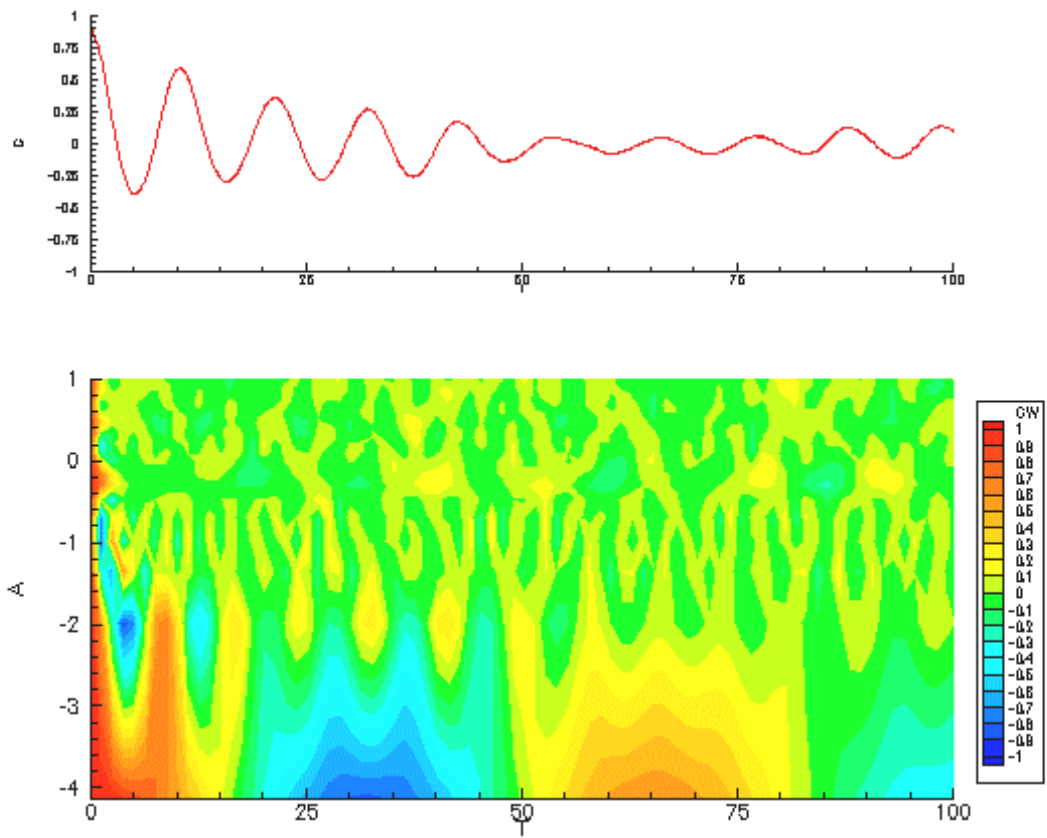

Figure 2 - Standard auto-correlation function (upper part) and local wavelet auto-correlation function (lower part)

On the other hand, when we consider the local wavelet auto-correlation function, we are able to distinguish the correlation properties for each scale (frequency). For example, if we are interested to know the auto-correlation characteristics for the main Schwabe cycle, it is sufficient to perform a cross-section of the wavelet map at the scale value around -2.39 (11 years). As we can see from Figure 2, the auto-correlation properties of the sunspot numbers are very different when viewed at different scales, and thus the standard auto-correlation function results not adequate. Considering in particular the Schwabe cycle as localised around 11 years and disregarding, for simplicity, its complex shape variation, we note that after 2 years the auto-correlation decays quickly to zero, reaching a peak of anti-correlation values at about 4 years, and only after the next 4 years we obtain again a significant increase of its value. Moreover, after 10-11 years the auto-correlation of the Schwabe cycle tends to very low average values. (Figure 3) This information can be very useful when we try to apply deterministic predictive methods for the main Schwabe cycle related to solar activity. It is useful to compare the cross-section of Figure 3 related to a constant scale with the average standard auto-correlation function.

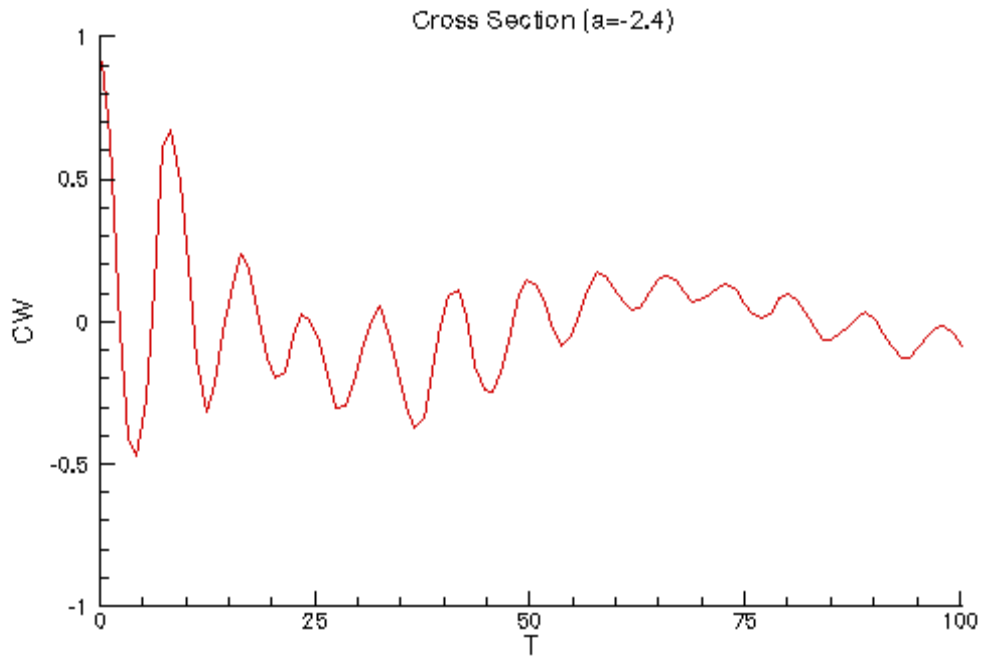

Figure 3 - Cross section of local wavelet auto-correlation map at 11 years periodicity

The U-shaped higher auto-correlation region limits the reliability of the solar cycle forecast to about 2 years, even if longer term prediction, limited from 6 to 10 years, seems to be again quite reliable. Medium term prediction, from 3 to 6 years and, in particular, very long term prediction beyond 11 years, appear more difficult, due to the loss of auto-correlation of sunspot numbers. In order to correctly evaluate these results it is important to stress that this analysis is limited to the linear framework of the auto-correlation functions.

*CONCLUSIONS*

Wavelet analysis allows a detailed investigation of the auto-correlation properties of a signal at different scales through a quantity, called local wavelet auto-correlation function. The application of this formalism to the temporal variations of solar activity, from the monthly time series of sunspot numbers, gives some interesting information about the correlation properties of different periodicities present in the signal and, in particular, the correlation characteristics of the main 11 years Schwabe cycle. The analysis of the wavelet auto-correlation map, allowed some interesting information about the range of reliable predictions reachable through deterministic predictive models.

*REFERENCES*

[1] Ochadlick, A. R., Kritikos, H.N., Giegengack, R. (1993)
    Geophys. Res. Lett. 20,14
[2] Lawrence, J.K., Cadavid, A.C., Ruzmaikin, A.A. (1995)
    ApJ, 455,366
[3] Sello, S. (2000)
    LANL Preprint Archive physics/0001042
[4] Tobias, S.M., Weiss, N.O., Kirk, V. (1995)
    Mont.Not.R.Astron. 502,273,1150
[5] Qin, Z. (1996)
    Astron. Astrophys. 310,646
[6] Sello, S. (1999)
    LANL Preprint Archive physics/9906035
[7] Torrence, C., Compo, G. P., (1998)
    Bull. Am. Mat. Soc. 79,1
[8] Foster, G. (1996)
    Astron. J. 112, 4
[9] Bendat, J.S., Piersol, A.G. (1971)
    Random Data: Analysis and measurement Procedures,
    Wiley Interscience (ed.)
[10] Forsth, M. (2000)
    in: Astrophysical Dynamics, Onsala Space Observatory
    A.B. Romeo (ed.)
[11] Sunspot Index Data Center Bruxelles
    http://www.oma.be/KSB-ORB/SIDC/index.html